\documentclass[11pt]{article}
\usepackage{amssymb,latexsym,amsmath,amsthm,graphicx,amscd,float}

\makeatletter \@addtoreset{figure}{section}
 \def\thefigure{\thesection.\@arabic\c@figure}
\def\fps@figure{h, t}
\@addtoreset{table}{bsection}
\def\thetable{\thesection.\@arabic\c@table}
\def\fps@table{h, t}
\newtheorem{theorem}{Theorem}[section]

\newtheorem{proposition}[theorem]{Proposition}

\newfont{\tenbi}{cmbxti10}
\newcommand{\la}{\lambda}

\DeclareMathOperator{\Prym}{Prym}
\DeclareMathOperator{\Jac}{Jac}

\DeclareMathOperator{\Cov}{Cov}
\begin{document}

\title{A shortcut to the Kovalevskaya curve \\ via pencils of genus 3 curves}

\author{Yu.N.Fedorov \\
Department of Mathematics, Polytechnic university of Catalonia, \\
Barcelona, Spain \\
 {\tt Yuri.Fedorov@upc.edu} \\ 
Luis C. Garc\'ia-Naranjo \\
Departamento de Matem\'aticas y Mec\'anica, IIMAS-UNAM, \\ Mexico City, Mexico \\
{\tt luis@mym.iimas.unam.mx} \\ 
Joan C. Naranjo \\
Departament d'Algebra i Geometria, Universitat de Barcelona \\
Barcelona, Spain \\
{\tt jcnaranjo@ub.edu} }

\maketitle

\begin{abstract}
We present a systematic way of derivation of the algebraic curves of separation of variables
for the classical Kovalevskaya top and its generalizations, starting from
the spectral curve of the corresponding Lax representation
found by Reyman and Semonov-Tian-Shansky.
 In particular, we show how the known Kovalevskaya curve of separation can be obtained,
by a simple one-step transformation, from the spectral curve.

The algorithm works for the general constants of motion of the system and is based on
W. Barth's description of Prym varieties via pencils of genus 3 curves.
It also allows us to derive new curves of separation of variables in various generalizations of this system.
 \end{abstract}

\section{Introduction}
The famous Kovalevskaya integrable case of heavy rigid body dynamics in $\mathbb R^3$
has various nontrivial integrable generalizations. One of the most interesting ones
is the motion of the Kovalevskaya
gyrostat in one or two constant force fields (e.g., gravitational and electric) 
described in \cite{RS, Yehia87, brs89}. The systems are Hamiltonian on the cotangent bundle $T^* SO(3)$ and,
as was shown in \cite{brs89, Yehia87}, their generic invariant manifolds are 3-dimensional tori.
 
In the present paper we consider only the cases of the presence of $SO(2)$-symmetries, 
which allow reductions to systems with two degrees of freedom with 2-dimensional generic
invariant tori. (The nature of the symmetries will be explained in Section 2.) 
This occurs in two situations:

1) the two fields are orthogonal and the intensities are equal; \\
2) there is just one constant force field, the corresponding system is
the Kovalevskaya gyrostat, which, for zero value of the gyrostatic momentum, 
becomes the classical Kovalevskaya top.

In \cite{brs89} Bobenko, Reyman and Semenov presented a Lax representation for both cases and gave
their algebraic geometric description. In particular, it was shown that, 
after factoring out a trivial involution, the corresponding spectral curve $C$ is non-hyperelliptic of
genus 3 and is a 2-fold ramified covering of an elliptic curve $E$.

The reduced systems were linearized on 2-dimensional Prym variety
$\Prym(C/E)\subset \Jac(C)$ associated with the covering $C \to E$. 
In the case of the classical Kovalevskaya top the authors of \cite{brs89}
demonstrated that the generic complex invariant manifolds are open subsets of $\Prym(C/E)$. 
This variety has polarization (1,2),
so it cannot be the Jacobian variety of a genus 2 curve which can be used
for a separation of variables of the systems.

On the other hand, in her celebrated paper \cite{kow}, S. Kovalevskaya presented a separation of variables
on a genus 2 (and, therefore, a hyperelliptic) curve $\cal K$,
whereas the complex flow of the system was linearized on the Jacobian of $\cal K$.

Apparently, the papers \cite{hvm89, avm88} were the first ones where this seeming 
contradiction had been explicitly explained:
the subvariety $\Prym(C/E)$ can be viewed as 2-fold covering of the Jacobians of 3 different, i.e., birationally
non-equivalent, genus 2 curves. In addition, there exist 3 other different genus 2 curves whose Jacobians are 2-fold coverings of
$\Prym(C/E)$. In \cite{hvm89} it was shown that, for the case of the Kovalevskaya top,
one of these hyperelliptic curves is birationally equivalent to the original curve of separation $\cal K$. (These relations
between $\Prym(C/E)$ and the Jacobians are briefly described in Section 3.)

Next, following a remarkable geometric construction of W. Barth \cite{bw85} inspired by \cite{hai83},
the papers \cite{hvm89, avm88} presented a family (pencil) $D_\lambda$, $\lambda \in {\mathbb P}$
of (generally non-hyperelliptic) genus 3 curves covering elliptic curves, 
and the coverings give rise to one and the same Prym variety {\it dual} to $\Prym(C/E)$ arising in the
Kovalevskaya problem. It was shown that all the above six
genus 2 curves can explicitly be extracted from $D_\lambda$, either as regularizations of singular
curves or as quotients of hyperelliptic curves of ${\cal D}_\la$.
Curiously, the spectral genus 3 curve $C$ of \cite{brs89} did not appear in this analysis.

The main observation of the present paper is that, in fact,
the spectral curve $C$ naturally fits into the W. Barth construction of pencils of genus 3 curves.
Using this property,
we show that the Kovalevskaya curve $\cal K$ can be obtained from $C$ by a simple one-step transformation,
even without knowing explicitly the family $D_\lambda$ and avoiding complicated calculations.
Namely, $\cal K$ appears as regularization of the curve {\it dual} to the spectral curve $C$ (see Section 3.)

In the particular case of zero area integral, the curve $C$ itself becomes singular of genus 2, then the above
construction implies that $\Jac(C), \Jac({\cal K})$ must be isogeneous, whereas $C,{\cal K}$  themselves are related
via the Richelot transformation. This property has previously been described in \cite{lm99}.

The same main observation allows us to recover easily the curves of separation of variables in all known linearizations
of the Kovalevskaya gyrostat in one and two fields 
(Section 4). We also derive some new separation curves for this problem (without doing the linearization itself).

\section{Equations of motion, spectral curves, and curves of separation}
In this section we reproduce some important known results on the dynamics and complex geometry of the Kovalevskaya gyrostat.

Using, for convenience, the notation of \cite{brs89},
consider the motion of an axisymetric rigid body in ${\mathbb R}^3$ with angular velocity vector
$\omega=(\omega_1,\omega_2, \omega_3)$
about a fixed point in presence of constant force fields $g, h\in {\mathbb R}^3$ of different nature 
(e.g., gravitational and electric), which are not parallel. 
Let $c_1, c_2$ be position vectors of the 
mass and, respectively, charge centers of the body. The equations of motion in the body frame have the form
\begin{equation} \label{ep0}
\begin{aligned}
\frac{dl}{dt} = [l,\omega] + [c_1,g] + [c_2,h], \\
\frac{dg}{dt} = [g,\omega], \quad \frac{dh}{dt} = [h,\omega],
\end{aligned}
\end{equation}
where $l=(l_1, l_2, l_3)$ is the total angular momentum and $[\cdot ,\cdot ]$ denotes the vector product.

Assume that the body carries a rotator with a fixed axis, then $l,\omega$ are related as
$\omega = {\cal J} l+ \kappa$, where ${\cal J}^{-1}$ is the inertia tensor of the body (with the rotator)
and $\kappa$ is related to the corresponding constant gyrostatic momentum. 

The phase space of the system is the cotangent bundle $T^* SO(3)$, 
and the components of $g,h$ stand for redundant coordinates on $SO(3)$. 
 Equivalently, the phase space can be regarded as dual to the Lie algebra
$so(3,2) = so(3)+ {\mathbb R}^3 + {\mathbb R}^3$, the semi-direct product of $so(3)$ and 
${\mathbb R}^3 \times {\mathbb R}^3$ with the corresponding Lie--Poisson structure. 
Then the system \eqref{ep0} is Hamiltonian with the Hamilton function
$$
 H= \frac 12 \langle {\cal J}l,l \rangle + \langle \kappa,l \rangle
- \langle g,c_1 \rangle -\langle h, c_2 \rangle ,
$$
which is the total conserved energy. Also, the intensities $|g|^2, |h|^2$ and the product
$\langle g,h \rangle$ are trivial integrals of the system \eqref{ep0}. They are Casimir functions of the
Lie--Poisson bracket of $so^*(3,2)$, and fixing generic values of the functions one obtains
a 6-dimensional coadjoint orbit in the coalgebra.

Assuming now ${\cal J}=\text{diag}(1,1,2)$, $c_1=(1,0,0), c_2=(0,1,0)$, and $\kappa=(0,0,\gamma)$ 
(the axes of symmetry
of the body and of the rotator are parallel),  we obtain the configuration of the
{\it Kovalevskaya gyrostat} with the corresponding Hamiltonian
\begin{equation} \label{H}
H= \frac 12 (l_1^2+l_2^2+2 l_3^2+2 \gamma l_3) -g_1 -h_2 \, . 
\end{equation}

For this case of equations \eqref{ep0}, in \cite{brs89} Bobenko, Reyman, and Semenov-Tyan-Shansky presented a
$4\times 4$ matrix Lax representation $\dot L(\la)=[L(\la), A(\la)]$ with a spectral parameter $\la\in {\mathbb P}$ and
\begin{gather} 
L(\la) = L_{-1} \la^{-1} + L_0 + L_1 \la,  \label{L_m01} \\
L_1 = \begin {pmatrix} 0&0&0&0\\ 
0&0&0&0\\ 0&0&-2&0 \\ 0&0&0&2\end{pmatrix} , \; 
L_0 = \begin {pmatrix} 
0 &-\gamma &-l_{2} &-l_{1} \\ 
\gamma & 0 & l_{1} &-l_{2} \\ 
l_{2} &-l_{1} & 0 &-2\,l_{3}-\gamma\\ 
l_{1} & l_{2} & 2\,l_{3} + \gamma & 0\end{pmatrix}, \notag \\  
L_{-1}= \begin{pmatrix} 
g_{1}-h_{2}& g_{2}+ h_{1} & g_{3}&h_{{3}}\\ 
g_{2}+h_{1} & h_{2}-g_{1} &h_{3} &-g_{{3}} \\ 
g_{{3}}&h_{{3}}&-g_{{1}}-h_{{2}}&g_{{2}}-h_{{1}}\\ 
h_{3} &-g_{{3}}&g_{{2}}-h_{{1}}&g_{{1}}+h_{{2}}
\end {pmatrix}  \notag
\end{gather} 
(we do not reproduce $A(\la)$ here) with the spectral curve 
$$
{\cal S} \, :\; |L(\la) - \mu {\bf I}|=0 \quad \text{of the form} \quad 
\mu^4- 2d_1(\la^2)\mu^2 + d_2(\la^2)=0,
$$
where $d_1(\la^2), d_2(\la^2)$ are Laurent polynomials.
 Apart from the Hamiltonian $H$ and the trivial integrals, the coefficients of $\cal S$
involve two other independent integrals $I_1=I_1(l,g,h), I_2=I_2(l,g,h)$ given by expressions (1.9), (1.10)
in \cite{brs89}, and all of them were shown to commute.
As a result, the Kovalevskaya gyrostat \eqref{ep0}, \eqref{H} is completely
integrable, and its generic real invariant varieties are 3-dimensional tori.

Next, the complex flow of the system was shown to be linear on the Jacobian variety of ${\cal S}$.
The latter has an obvious involution $\tau_1 \,:\, (\la,\mu) \to (-\la, \mu)$,
so it is a 2-fold covering of the quotient curve
\begin{equation} \label{gen_C}
   C = {\cal S} /\tau_1 \,:\; \mu^4- 2d_1(z)\mu^2 + d_2(z)=0
\end{equation}
of maximal genus 4.
It follows that the Jacobian variety of ${\cal S}$, $\Jac({\cal S})$ contains $\Jac(C)$ as an Abelian
subvariety.
As was shown in \cite{brs89}, the flow is confined to the subtorus $\Jac(C)\subset \Jac({\cal S})$.
For this reason, it is sufficient to study only the properties of $C$, which
will be called the quotient spectral curve.

\paragraph{The cases of reduction and the corresponding spectral curves.} 
As was first shown in \cite{BorMam, Yehia87}, when the force fields are orthogonal of equal intensities, that is,
$|g|=|h|={\bf a}, \langle g,h \rangle =0$, the system possesses
an additional $SO(2)$-symmetry group consisting of simultaneous rotations about the symmetry 
axis of the gyrostat and the fixed in space vector $[g,h]$. The group action
is described by the Hamiltonian $F_2= l_3- \langle l, [g,h]\rangle/|g|^2$, which is a conserved
quantity. 


Let ${\bf R}= {\bf a}^{-1} (g\, h\, [g,h])^t \in SO(3)$ be the rotation matrix of the gyrostat. Following 
\cite{BorMam,Harl_Yehia87}, the action of $SO(2)$ on $T^*SO(3)$ is generated by conjugations
\begin{equation} \label{action_R}
 {\bf R} \mapsto   g_\tau {\bf R} g_\tau^{-1}, \quad
 g_\tau = \begin{pmatrix} \cos\tau & \sin\tau & 0 \\
  -\sin\tau & \cos\tau & 0 \\ 0 & 0& 1 
\end{pmatrix} , \quad \tau \in [0,2\pi). 
\end{equation}
 
As was shown in \cite{BorMam, BorMam_book}, in the quaternion realization of the group $SO(3)$ as the
factor of $S^3=\{ \la_0^2 + \la_1^2+\la_2^2+ \la_3^2=1 \}$ by ${\mathbb Z}^2$, the $SO(2)$-symmetry acts
as the rotation in the 2-plane $(\la_1,\la_2)$, then the quotient variety $SO(3)/SO(2)$ is the 
until sphere $S^2=\{ \la_0^2 + \la_3^2 + s^2=1 \}$ with $s=\sqrt{\la_1^2+\la_2^2}$. 

For any constant $c\in {\mathbb R}$, the level variety $F_2^{-1}(c) \in T^*SO(3)$ is 5-dimensional,
and the quotient space $F_2^{-1}(c)/SO(2)$ is isomorphic to a globally nontrivial fiber bundle $T\, S^2$.     
 
Following \cite{BorMam}, the system \eqref{ep0}, \eqref{H} admits a Hamiltonian reduction onto 
$T S^2$ endowed with the standard symplectic structure of $T^* S^2$ plus a singular magnetic term.   
The reduced gyrostat in 2 fields possesses two independent commuting integrals, 
which are reductions of $H, I_2$ (or $I_1, I_2)$, and is linearized on 2-dimensional
invariant tori.     
 
In the sequel we will set $|g|=|h|=1$, then the integrals $I_1, I_2$ read 
\begin{align}
 I_1 & = 2H + F_2^2 + 2\gamma F_2,  \qquad  \label{I_F} \\
I_2 & = (l_1^2 -l_2^2 +2g_1 -2h_2)^2 + 4(l_1 l_2+g_2+h_1)^2 \notag \\
& \qquad \qquad - 4\gamma \left( (l_3+\gamma)(l_1^2+l_2^2) +2l_1 g_3+ 2l_2 h_3 \right),  \label{I2}
\end{align}
where the value of the momentum $F_2$ should be fixed. 

From Section 5 of \cite{brs89} we extract the following expression for the quotient curve $C$
\begin{gather}
C=C_2 \, :\;  \mu^4- 2d_1(z)\mu^2 + d_2(z)=0, \label{Cs} \\
d_1(z) = \frac 2z - (2 H+ \gamma^2) + 2z, \quad
d_2(z)= 4\frac{I_1- 2H-\gamma^2}{z} + (I_2 +4 \gamma^2 H + \gamma^4)-4 \gamma^2 \, z , \notag
\end{gather}
which, under the birational transformation $z=1/x, \mu= y/x$, can be written as 
\begin{equation} \label{sp_gen}
 C_2\, :\;  y^2 = (2 x^2-(2H+\gamma^2)x+2) x + 2 x \sqrt{ x^4 -I_1 \, x^3 +(H^2-I_2/4+2)x^2 - 2 Hx+ 1 }\, .
\end{equation}
The curve has the involution $\sigma\, :\, (x,y) \to(x,-y)$, hence
it is a 2-fold covering of the elliptic curve $E_2$ given by equation
\begin{equation} \label{E22}
E_2 \, : \quad y^2 =x^4 -I_1 x^3 +(H^2-I_2/4+2)x^2 - 2 Hx+ 1.
\end{equation}
The covering is ramified over 4 points on $E_2$, therefore,
by the Riemann--Hurwitz formula, for generic values of the constants of motion, genus($C_2)=3$.
As we will see in Section 4,
$C_2$ is hyperelliptic (i.e., can be transformed to a hyperelliptic form), this quite special property
will simplify our analysis considerably.
\medskip

In the other case of $SO(2)$-symmetry, the Kovalevskaya gyrostat moves only
in one constant force field (one can set $h=0$), and for $\gamma=0$ the system becomes the
classical Kovalevskaya top. The $SO(2)$-action is the rotation about the fixed vector $g$ and is 
generated by the Hamiltonian $F_1= \langle l, g \rangle$, i.e., the area integral. 
Upon fixing its value, the system admits a Hamiltonian 
reduction onto the cotangent bundle $T^*S^2$ of $S^2=\{ \langle g, g \rangle =1 \}$ with 
two independent commuting integrals and 2-dimensional generic invariant tori.

Equivalently, the reduced Kovalevskaya gyrostat can be regarded as a Hamiltonian system on the coalgebra
$e^*(3)={\mathbb R}^6 (l,g)$ with the two integrals $H(l,g), I_2(l,g)$ 
and the Casimir functions $F_1, \langle g, g \rangle$.   
 
In the considered case the quotient spectral curve $C$ takes the following concrete form
\begin{gather}
 C=C_1 \, :\;  \mu^4- 2d_1(z)\mu^2 + d_2(z)=0, \label{C0} \\
d_1 (z) = \frac 1z - (2 H+ \gamma^2) + 2z, \quad d_2(z) = \frac{1}{z^2} + 4\frac{I_1- H-\gamma^2/2}{z} +
(I_2 +4 \gamma^2 H + \gamma^4) -4 \gamma^2 \, z , \notag
\end{gather}
where now $I_1 = F_1^2= \langle l, g \rangle^2$, and $I_2$ is obtained from \eqref{I2}
by setting $h=0$.
Under the same birational transformation $z=1/x, \mu= y/x$ the curve $C_1$ can be written in the form
\begin{equation} \label{sp_kow}
  C_1\, :\;  y^2 = (x^2-(2H+\gamma^2) x+2) x + x \sqrt{ -I_1 \, x^3 +(H^2-I_2 /4+1)x^2 - 2 Hx+ 1 }\, .
\end{equation}
The factor of $C_1$ by the involution $\sigma\, :\, (x,y) \to(x,-y)$ is the elliptic curve
$$
  E_1 \, :\; Y^2 = -I_1 x^3 +(H^2-I_2 /4+1)x^2 - 2 Hx+ 1 ,
$$
and again, generally (also for $\gamma =0$), genus$(C_1)=3$.
Note that, in contrast to $C_2$, the curve $C_1$ is not hyperelliptic, even for $\gamma =0$.
\medskip

In both considered cases the involution $\sigma$ on $C$ extends to its Jacobian.
Thus the latter contains two Abelian subvarieties: the elliptic curve $E=E_1$ or $E_2$ 
itself and the 2-dimensional
Prym variety, $\Prym(C/E)$, which is anti-symmetric with respect to the extended involution, whereas
$E$ is invariant. Equivalently, following \cite{Mum}, $\Prym(C/E)=\text{ker}\, (1+\sigma)$.

Using the approach of vector Baker--Akhiezer functions, the authors of \cite{brs89} linearized 
the complex flow of the reduced gyrostat in one field on the subvariety $\Prym(C_1/E_1)\subset \Jac(C_1)$ 
and proved that, for $\gamma=0$ and $F_1\ne 0$, the generic 2-dimensional complex
invariant manifold $\cal I$ of the system, the common level surface of the integrals 
$H, F_1, I_2, \langle g, g \rangle$, is an open subset of the Prym subvariety.

It is known (see \cite{fay73} and Section 3 below) that
this variety has polarization (1,2), so it cannot be the Jacobian of a genus 2 curve which can be used
for a separation of variables of the system.

One can compare this result with the famous Kovalevskaya reduction of the problem to
hyperelliptic quadratures\footnote{Details of this reduction can be found in many publications, we
do not reproduce them here.}
$$
  \frac{d s_1}{w_1} + \frac{d s_2}{w_2}=0, \quad  \frac{s_1\, ds_1}{w_1} + \frac{s_2\, ds_2}{w_2}
= \sqrt{-1}\, dt \, ,
$$
where the pairs $(s_1,w_1), (s_2,w_2)$ are coordinates of two points on the genus 2
Kovalevskaya's separation curve:
\begin{equation} \label{KC}
{\cal K}\, :\;  w^2 = [s \left( (s-H)^2 +1 -I_2/4 \right)- I_1] \left( (s-H)^2-I_2/4 \right) .
\end{equation}
The integral form of the quadratures defines the Abel--Jacobi map $\text{Sym}^2 {\cal K} \mapsto \Jac({\cal K})$, which
suggests that the complex invariant tori $\cal I$ of the system and $\Jac({\cal K})$ must be algebraically related.

This relation was first described in \cite{hvm89, avm88}: Let $\hat {\cal I}$ be a compactification of $\cal I$ converting
it to an Abelian torus, then $\hat {\cal I}$ is an 8-fold unramified covering on $\Jac({\cal K})$.
Equivalently, if we replace $\Jac({\cal K})$ by its 16-fold
covering, obtained by doubling all the 4 period vectors (both Abelian tori are conformally equivalent),
then $\Jac({\cal K})$ can be regarded as a 2-fold unramified covering of $\hat{\cal I}$.

Note that the detailed algebraic geometric analysis of \cite{hvm89, avm88} did not involve neither the Lax pair, nor the
quotient spectral curve $C_1$ in \eqref{C0}, but a pencil of genus 3 curves covering elliptic curves and
leading to the same Prym variety, which was identified with the manifold $\hat{\cal I}$.

\paragraph{Remark 1.}
As was noticed in \cite{brs89}, for the case of zero area integral ($I_1=0$)
the quotient spectral curve $C_1$ of the Kovalevskaya top
(the gyrostatic component $\gamma$ is zero) is singular, and its regularization is a genus 2 curve
$C_{10}$ with hyperelliptic form\footnote{One can verify that in the case $I_1=0$
the curve $C_1$ is singular of genus 2 also for $\gamma\ne 0$.}
\begin{equation} \label{C_10}
    W^2 = Z (Z^2 +2H Z+ I_2/4) (Z^2+2 H Z+ I_2/4-1) ,
\end{equation}
which can be obtained from \eqref{sp_kow} by the birational change of coordinates
$$
   Z= \frac{y^2-x^3}{2 x^3}, \quad W= y \frac{x^3-y^2}{\sqrt{2}\, x^4} .
$$
In Section 7.6 of \cite{brs89} it was also claimed that in this case
$\Prym(C_1/E_1)$ coincides with $\Jac(C_{10})$ and that the complex invariant manifold ${\cal I}$ 
is an open subset of a 2-fold unramified covering of $\Jac(C_{10})$.  
In fact, a careful analysis based on the
procedure described in \cite{be77} shows that $\Prym(C_1/E_1)$ can be recovered as an
Abelian subvariety of the 3-dimensional generalized Jacobian of the singular curve $C_1$, which is
an extension of $\Jac(C_{10})$ by ${\mathbb C}^*$.
Then $\Prym(C_1/E_1)$ appears as a smooth 2-fold covering of $\Jac(C_{10})$.
As a result, ${\cal I}$ is an open subset of $\Prym(C_1/E_1)$ in both cases: $F_1\ne 0$ and $F_1=0$.      
Moreover, the proof of this statement equally holds for $\gamma \ne 0$. Thus we have
 
\begin{theorem} \label{Prym-tori}
For general constants of motion, including $F_1=0$, the compactified 2-dimensional complex
invariant manifold $\hat{\cal I}$ of the reduced Kovalevskaya gyrostat in one field is $\Prym(C_1/E_1)$.
\end{theorem}

Next, when performing a linearization of the complex flow on $\Jac(C_{10})$ in the case $F_1=0$,
Bobenko, Reyman, and Semenov \cite{brs89} observed that $C_{10}$ is not birationally equivalent to the
Kovalevskaya curve \eqref{KC} with $\gamma=0$.
A decade later Lepr\"ovost and Markushevich \cite{lm99} proved that the Jacobians of both curves
are isogeneous:
$\Jac({\cal K})$ is a 4-fold unramified covering of $\Jac(C_{10})$,
obtained by duplicating two of the 4 period vectors
of $\Jac(C_{10})$ (namely, by duplicating the Riemann period matrix of $\Jac(C_{10})$).
The curves $C_{10}, {\cal K}$ themselves are related algebraically via the Richelot transformation, whose
description can be found in \cite{bm88,lm99}.

Below in Section 4 we will see that the Richelot relation between $C_{10}, {\cal K}$ becomes quite natural
in the context of pencils of genus 3 curves and dual
Prym varieties. This global construction described in the next section
will also allow us to recover all the genus 2 curves that previously
appeared in the linearization of the Kovalevskaya gyrostat in one and two force fields,
as well as to construct new genus 2 curves of separation of variables.

Apart from the mentioned generalizations, there are also various deformations of
the Kovalevskaya gyrostat obtained by modifications of the Poisson bracket, see e.g., \cite{st02}.
We do not consider here the corresponding curves of separations.

We conclude the section with the following analog of Theorem \ref{Prym-tori} for the Kovalevskaya gyrostat in two fields.

\begin{theorem} \label{Prym-tori2}
For general constants of motion, the complex 2-dimensional invariant manifold ${\cal I}$ of the reduced Kovalevskaya gyrostat in two fields is an open subset of $\Prym(C_2/E_2)$.
\end{theorem}

\noindent{\it Proof.}
Let ${\cal I}_{C_2} \subset T^*SO(3)$ be the complex 3-dimensional invariant manifold of the gyrostat in two fields 
obtained by fixing the independent constants of motion $H, I_1, I_2$ and, therefore, the quotient spectral curve $C_2$ in \eqref{Cs}. 
 
Obviously, ${\cal I}_{C_2}$ is isomorphic to the complex isospectral manifold ${\cal J}_{C_2}$, 
the set of all the matrix Laurent polynomials $L(\la)$ of the form \eqref{L_m01} having the same spectral curve $\cal S$ and its quotient $C_2$. Consider {\it the eigenvector map}
$$
{\cal M}\, :\;  {\cal J}_{C_2} \longrightarrow \Jac({\cal S}) ,
$$
defined in the following standard way:  a matrix $L(\la)\in {\cal J}_{C_2}$ induces
the eigenvector bundle ${\mathbb P}^2\to {\cal S}$: for any point $p=(\la,\mu)\in S$
$$
p \longrightarrow \psi (p) = (\psi_1(p), \dots, \psi_4(p))^T \quad \text{such that} \quad
L(\la) \psi(p) = \mu\, \psi(p) .
$$
We assume that the eigenvector $\psi(p)$ is normalized: $\langle \alpha, \psi(p)\rangle \equiv 1$, for
a certain $\alpha \in {\mathbb P}^3$. This defines the divisor $\cal D$ of poles of $\psi (P)$ on $\cal S$.
For any choice of normalization, such divisors form an equivalence class $\{ {\cal D} \}$.  For a base point $p_0\in {\cal S}$,
the class defines a point  ${\cal D}- N p_0 \in \Jac(S)$, where $N=\text{degree}({\cal D})$. 
Then one has ${\cal M}(L(\la))= {\cal D}- N p_0$.   

In Lemma 4.5 of \cite{brs89} it was shown that under the time evolution of $L(\la)$ the image ${\cal M}(L(\la))$ is restricted to a translation of
$$
  \Prym(C_2/E_2) \subset \Jac(C_2) \subset \Jac({\cal S}) .
$$ 

Let $\mathfrak G$ be the maximal subgroup of ${\mathbb P}GL_4({\mathbb C})$ which acts freely  on ${\cal J}_{C_2}$ by conjugations and preserves the form of $L(\la)$. 
For any $g\in {\mathfrak G}$, the $\cal M$-images of $L(\la)$ and $\tilde L(\la)= g L(\la) g^{-1}$ coincide. Then the eigenvector map pushes down to 
$$
{\cal M}'\, :\;  {\cal J}_{C_2} / {\mathfrak G}  \longrightarrow   \Prym(C_2/E_2) ,
$$
and, as was shown in e.g., \cite{AvM80, Beau90}, such a map is {\it injective}.

Now note that $\mathfrak G$ must be a stabilizer of $L_1$ in  \eqref{L_m01}, as well as of the constant $2\times 2$  block 
$\begin{pmatrix} 0 & -\gamma \\ \gamma & 0 \end{pmatrix}$ of $L_0$. It also must preserve the symmetry of $L_{-1}$. Then, nesesarily,  any element of $\mathfrak G$
has the block form 
$$
\begin{pmatrix} r &  {\bf 0} \\  {\bf 0} & {\bf I} \end{pmatrix}, \quad r \in SO(2,{\mathbb C}) , \quad {\bf I} = \begin{pmatrix} 1 & 0 \\ 0 & 1 \end{pmatrix} .
$$

Since both varieties ${\cal J}_{C_2} / {\mathfrak G}, \Prym(C_2/E_2)$ have dimension 2 and ${\cal M}'$ is injective, 
the factor ${\cal J}_{C_2} / {\mathfrak G}$ is an open subset of $\Prym(C_2/E_2)$. 

We now show that the action of  ${\mathfrak G}$ is generated by the action of $SO(2,{\mathbb C})$ on $T^*SO(3)$ induced
 by \eqref{action_R} with $\tau\in {\mathbb C}$. In view of the expressions for $g,h,l$ in terms of Euler angles and their time derivatives,  \eqref{action_R} implies the transformation $(g,h,l)\to (\tilde g,\tilde h,\tilde l)$ with 
\begin{align*}
 2\tilde g_{1} &=\cos \left( 2\,\tau \right)  \left( g_{1}-h_{2} \right) -\sin \left( 2\,\tau \right)  \left( g_{{2}}+h_{{1}} \right) +g_{{1}}+h_{2}, \\
2\tilde g_{2} & =\cos \left( 2\,\tau \right)  \left( g_{2}+h_{1} \right) +\sin \left( 2\,\tau \right)  \left( g_{{1}}-h_{{2}} \right) +g_{{2}}- h_{1},\\ 
\tilde g_{3} & =\cos \left( \tau \right) g_{{3}}-\sin \left( \tau \right) h_{3},\\ 
2\tilde h_{1} &=\cos \left( 2\,\tau \right)  \left( g_{{2}}+h_{{1}} \right) -\sin \left( 2\,\tau \right)  \left( h_{{2}}-g_{{1}} \right) -g_{{2}}+h_{{1}},\\ 
2\tilde h_{2} &=\cos \left( 2\,\tau \right)  \left( h_{{2}}-g_{{1}} \right) +\sin \left( 2\,\tau \right)  \left( g_{{2}}+h_{{1}} \right) +h_{{2}}+g_{{1}},\\ 
\tilde h_{3} &=\cos \left( \tau \right) h_{3}+\sin \left( \tau \right) g_{3} , \\
\tilde l_1 &= \cos (\tau) l_1-\sin(\tau) l_2, \quad \tilde l_2= \sin (\tau) l_1+\cos(\tau) l_2, \quad 
\tilde l_3 =l_3 \, .
\end{align*}
The latter can be described as the conjugation
$$
 L(\tilde g,\tilde h,\tilde l ; \la) = g(\tau)   L( g, h, l ; \la) g^{-1}(\tau), \quad g = \begin{pmatrix}  \cos \tau & -\sin \tau & 0 & 0 \\
																				 \sin \tau & \cos \tau & 0 & 0 \\
																					0 & 0 & 1 & 0 \\ 0&0&0&1 \end{pmatrix} .
$$
As a result, the action of  ${\mathfrak G}$ on  ${\cal J}_{C_2}$ coincides with that of   $SO(2,{\mathbb C})$ on ${\cal I}_{C_2}$.
Then ${\cal I}_{C_2}/SO(2,{\mathbb C})$, which is the complex invariant manifold of the reduced gyrostat in 2 fields, 
is isomorphic to  ${\cal J}_{C_2} / {\mathfrak G}$, which, in turn, is isomorphic to an open subset of $\Prym(C_2/E_2)$. 
$\square$

\section{Dual (1,2)-$\Prym$ varieties, isogeneous hyperelliptic Jacobians, and dual pencils of genus 3 curves}

We start this section with the description of the general form of genus 3 curves (over the field $\mathbb C$)
covering elliptic curves and the related Prym varieties.

\begin{theorem} \label{gen_cover} Let $E$ be a generic elliptic curve in ${\mathbb P}^2$ whose affine part in
${\mathbb C}^2(x,y)$ is given by equation
$$
y^2= \Phi(x): = (x-c_1)(x-c_2)(x-c_3).
$$
\begin{description}
\item{1)} Any 2-fold covering
$C\to E$ ramified at 4 arbitrary chosen finite points \\ $Q_1=(x_1,y_1), \dots, Q_4=(x_4,y_4)$ on
$E$ can be written in form
\begin{equation} \label{gen_gen_3}
C \; : \quad  w^2 = g_3(x) + (\alpha x+ \beta) y, \quad  y^2=\Phi(x),
\end{equation}
where $g_3(x)$ is a polynomial of degree at most 3 and $\alpha,\beta$ are constants such that
\begin{equation} \label{branch_cond}
 \Psi(x): = g_3^2(x)- (\alpha x+ \beta)^2 \Phi (x) = (x-x_1)\cdots (x-x_4) \rho^2(x)
\end{equation}
for a certain linear function $\rho(x)$ (it is a constant if degree of $g_3(x)<3$ and $\alpha=0$).

\item{2)} For any divisor $Q=\{Q_1,\dots,Q_4\}$ on $E$, the set $\Cov_E^2(Q)$ of 2-fold covers of $E$ ramified
exactly at $Q$ consists of $2^2=4$ birationally non-equivalent covering curves $C$.
\end{description}
\end{theorem}

The idea of the proof of item (1) of the theorem is due to A. Levin \cite{LevA}.
For reasons of brevity we do not give it here, just mention that the condition \eqref{branch_cond} says that
on the curve $E$ the meromorphic function $w^2$ has 4 simple zeros, one double zero (given by the zero of $\rho(x)$)
and a pole of order 6. Multiple zeros or poles of $w^2$ of even order give singularities of $C$, which, after
the regularization, do not contribute to branching of $C \to E$. As a result
this covering has precisely 4 ramification points.

Note that the quotient spectral curves $C_{1}, C_2$  in \eqref{sp_kow}, \eqref{Cs} have the structure \eqref{gen_gen_3}.

The involution $\sigma : (x,w)\to (x,-w)$ on $C$ extends to $\Jac (C)$, which then
contains two Abelian subvarieties: the elliptic curve $E$ itself and the 2-dimensional
Prym variety denoted as $ \Prym(C/E)$, the latter being anti-symmetric with respect to $\sigma$, whereas
$E$ is invariant. The complex subtori $E,\Prym(C/E) \subset \Jac(C)$ intersect at 4 points,
which are the half-periods of $E$.

As an Abelian variety, $\Prym(C/E)$ is a complex torus ${\mathbb C}^2/{\Lambda}$,
with ${\Lambda}$ being a lattice generated by 4 period vectors.
In appropriate coordinates $z_1,z_2$ in ${\mathbb C}^2$, its period matrix takes the form
\begin{equation} \label{Lambda_periods}
\Lambda =  \left(
{\begin{array}{rrcc}
1 & 0 & a & b \\
0 & 2 & b & c
\end{array}}
 \right), \quad \text{equivalent to} \quad
\widehat \Lambda =\left(
{\begin{array}{rrcc}
2 & 0 & c & b \\
0 & 1 & c & a \end{array}}
 \right) .
\end{equation}
The right half of $\Lambda$ is the Riemann matrix $\tau=\begin{pmatrix}
 a & b \\
 b & c \end{pmatrix}$ with Im $\tau\, >0$, and
the diagonal of the left half indicates the polarization of $\Prym(C/E)$, namely (1,2).

According to \cite{bw85}, any 2-dimensional Abelian variety $\cal A$ with polarization (1,2) can be
realized as the Prym variety of a covering $C\to E$ described in Theorem \ref{gen_cover}.
Moreover, for a generic $\cal A$ there is a one-parametric family of 2-fold coverings $C_\la\to E_\la$,
in which both curves $C$ and $E$ depend on the parameter $\la \in {\mathbb P}$ (see also  below).

It is also natural to consider the (1,2)-polarized variety $\Prym^*(C/E)$
{\it dual}\footnote{A coordinate-free definition of the dual Prym variety can be found, e.g., in \cite{hai83} p. 466} to
$\Prym (C/E)$, its period matrix can be written
as
\begin{equation}
\label{Lambda*}
\Lambda^*=\left(
{\begin{array}{rrcc}
1 & 0 & c/2 & b  \\
0 & 2 & b  & 2a
\end{array}}
 \right).\end{equation}
This matrix is obtained by dividing by 2 the 1st and 3rd period vectors of  $\widehat\Lambda$
and rescaling $(z_1,z_2)\to (z_1, z_2/2)$ to get a Riemann matrix. Namely,
$$
\widehat\Lambda= \left(
{\begin{array}{rrcc}
2 & 0 & c & b \\
0 & 1 & b & a
\end{array}}
 \right) \to \;
\left(
{\begin{array}{rrcc}
1 & 0 & c/2 & b  \\
0 & 1 & b/2 & a
\end{array}}
 \right) \to \;
\left(
{\begin{array}{rrcc}
1 & 0 & c/2 & b  \\
0 & 2 & b  & 2\,a
\end{array}}
 \right).
$$
 Note that, modulo the conformal equivalence,
the dual to $\Prym^*(C,\sigma)$ is again $\Prym (C,\sigma)$, and
both varieties can be regarded as 4-fold coverings of each other.

\paragraph{Prym varieties and isogeneous Jacobians.}
As was mentioned in Section 6 of \cite{hvm89}, given an arbitrary (1,2)-polarized 2-dimensional Abelian variety
${\mathcal A}$ having the period matrix $\Lambda$ as in \eqref{Lambda_periods},
there are several ways to obtain a principally polarized 2-dimensional
Abelian variety (and therefore, the Jacobian of a genus 2 hyperelliptic curve) isogeneous to
${\mathcal A}$.
An obvious way is to divide the second period vector of $\Lambda$ by 2, to get the period matrix
$({\bf I},\tau)$, ${\bf I}=\text{diag}(1,1)$.
On the other hand, one can double the first period vector of $\Lambda$ to obtain the matrix
$(2{\bf I},\tau)$. The latter defines a torus with polarization (2,2),
which is conformally equivalent to a principally polarized variety with the period matrix
$\left({\bf I},\frac 12 \tau \right)$.

There are others, less obvious, transformations which produce Jacobian varieties
symplectically non-equivalent to the above ones (their explicit description is given in
\cite{hvm89}, \cite{EF15}). There is, however, a very limited number of them.

\begin{theorem}[ \cite{hvm89}, \cite{EF15} ] \label{main_th}
 For a generic covering $C\to E$ described in Theorem \ref{gen_cover}
there are exactly 6 birationally non-equivalent genus 2 curves $\Gamma_1, \Gamma_2, \Gamma_3$ and
$\tilde\Gamma_1, \tilde\Gamma_2, \tilde\Gamma_3$ such that
$\Prym(C/E)$ is a 2-fold unramified covering of $\Jac(\Gamma_{1}), \Jac(\Gamma_{2}), \Jac(\Gamma_{3})$, and
$\Jac(\widetilde\Gamma_{1}), \Jac(\widetilde\Gamma_{2}), \Jac(\widetilde\Gamma_{3})$ are
2-fold unramified coverings of $\Prym(C/E)$.  \\
Next, $\Jac(\Gamma_{1}), \Jac(\Gamma_{2}), \Jac(\Gamma_{3})$ are also 2 fold coverings of the dual variety
$\Prym^*(C/E)$,
whereas the latter is a 2-fold covering of tori conformally equivalent to
$\Jac(\widetilde\Gamma_{1})$, $\Jac(\widetilde\Gamma_{2})$, $\Jac(\widetilde\Gamma_{3})$
as depicted in the following diagram, where the arrows denote the corresponding 2:1 coverings and double
arrows indicate 4-fold coverings:
\begin{equation} \label{diagram}
\begin{array}{ccccc}
\Jac(\widetilde\Gamma_{1}) &  & \Jac(\widetilde\Gamma_{2}) &  & \Jac(\widetilde\Gamma_{3}) \\
\Vert & \searrow  & \downarrow  & \swarrow  & \Vert \\
\Vert &  &  \Prym(C/E ) &  & \Vert \\
\Downarrow  & \swarrow  & \downarrow & \searrow  & \Downarrow  \\
\Jac(\Gamma_1) &  & \Jac(\Gamma_{2}) &  & \Jac(\Gamma_3 ) \\
\Vert & \searrow  & \downarrow  & \swarrow  &  \Vert \\
\Vert &  & \Prym^* (C/E) &  &  \Vert \\
\Downarrow & \swarrow  & \downarrow  & \searrow  &  \Downarrow \\
\Jac(\widetilde\Gamma_{1}) &  & \Jac(\widetilde\Gamma_{2}) &  & \Jac (\widetilde\Gamma_{3})
\end{array}
\end{equation}
\end{theorem}

Below we identify conformally equivalent Abelian varieties obtained one
from another by duplication of all of the 4 periods. Under this convention,
$\Prym(C/E )$ can be regarded as an 8-fold covering of any of
$\Jac(\widetilde\Gamma_{1})$, $\Jac(\widetilde\Gamma_{2})$, $\Jac(\widetilde\Gamma_{3})$.

\paragraph{The Richelot correspondences.}
As was shown in \cite{EF15}, the period matrices of some of the Jacobians ''below'' and ''above''
$\Prym(C,\sigma)$ are obtained from one another by doubling their Riemann matrix $\tau$.
Then, upon a proper ordering of the Jacobians, the above diagram can be accompanied with the following one
\begin{equation} \label{tau->2tau}
\begin{array}{ccccc}
\Jac(\tilde \Gamma_1) &  & \Jac(\tilde \Gamma_2) &  & \Jac(\tilde \Gamma_3) \\
\uparrow  & \nwarrow  &  & \searrow  & \downarrow  \\
\Jac(\Gamma_1) &  & \Jac(\Gamma_2) &  & \Jac(\Gamma_3)
\end{array}
\end{equation}
where arrows denote the duplication $({\bf I}\, \tau) \to ({\bf I}\, 2\tau)$). Explicit relations between the
Riemann matrices of the Jacobians are given in \cite{EF15}.
The corresponding curves $\Gamma_\alpha, \widetilde\Gamma_\beta$ are related algebraically via the
{\it Richelot transformation} (explicit formulas can be found in \cite{bm88, lm99}).
One should stress that such a transformation is only a correspondence:
for example, $ \Jac(\tilde \Gamma_1)$ is obtained by duplication of symplectically non-equivalent Riemann matrices
of $\Jac(\Gamma_1)$ and $\Jac(\Gamma_{2})$.

\paragraph{The case of two pairs of elliptic curves.}
As was noted in \cite{hvm89}, Theorem \ref{main_th} does not completely hold in special cases when
$\Prym(C/E)$ contains Abelian subtori. In particular, this happens when one of the associated curves
in \eqref{diagram} say
$\tilde \Gamma_3$, is a 2-fold covering of two elliptic curves $E_-, E_+$ with normalized periods
$(1,\tau_-), (1, \tau_+)$ respectively.
Then, according to the Weierstrass--Poincar\'e theory of reduction (see e.g., \cite{Bel, fay73}),
$\Jac(\tilde \Gamma_3)$ contains
the curves $E_-, E_+$, although $\Jac(\tilde \Gamma_3)$ is not the direct product $E_- \times E_+$, but
there is 4-fold isogeny $\begin{CD}
E_+\times E_- @ > 4:1 >> \Jac(\tilde \Gamma_3) \end{CD}$. Then, according to \cite{EF15},
Theorem \ref{main_th} must be replaced by the following

\begin{proposition} \label{ell_case} Let $\Prym(C/E)$ be a 2-fold unramified covering of the direct product
$E_- \times E_+$. Then there are two non-equivalent genus 2 curves $\tilde\Gamma_2, \tilde\Gamma_3$, both
being 2-fold ramified coverings of $E_-, E_+$, such that
$\Jac(\widetilde\Gamma_{2}), \Jac(\widetilde\Gamma_{3})$ are 2-fold unramified coverings of $\Prym(C/E)$.

Next, $\Prym(C/E)$ is a 2-fold unramified coverings of Jacobians of other non-equivalent genus 2 curves
$\Gamma_1, \Gamma_2$, both being 2-fold ramified coverings of elliptic curves ${\cal E}_-, {\cal E}_+$ with
normalized periods $(1,2\tau_-), (1,\tau_+/2)$ respectively. Finally, the direct product
${\cal E}_- \times {\cal E}_+$ is a 2-fold unramified covering of $\Prym(C/E)$.
\end{proposition}

The relations between the genus 2 curves, their Jacobians,
and the elliptic curves are described by the following diagrams
(on the first one the arrows denote 4-fold coverings such that their inversions describe the Richelot
transformations).
\begin{gather}
 \begin{array}{ccccc}
{\cal E}_- \times {\cal E}_+ &  & \Jac(\tilde \Gamma_2) &  & \Jac(\tilde \Gamma_3) \\
\downarrow  & \searrow  & & \nwarrow   & \uparrow  \\
\Jac(\Gamma_1) &  & \Jac(\Gamma_2) &  & E_- \times E_+
\end{array}  \label{special_Jacs}  \\
\begin{CD}
E_- @ < 2:1 << \tilde\Gamma_2, \tilde\Gamma_3  @ > 2:1 >> E_+ \end{CD}, \quad
\begin{CD} {\cal E}_- @ < 2:1 << \Gamma_1, \Gamma_2  @ > 2:1 >> {\cal E}_+ \end{CD}.
 \end{gather}
Here the products ${\cal E}_- \times {\cal E}_+$, $E_- \times E_+$ replace $\Jac(\tilde\Gamma_1)$ and
$\Jac(\Gamma_3)$ in diagrams \eqref{diagram}, \eqref{tau->2tau} for the general case.

We stress that, as above, all the Abelian varieties are considered modulo conformal equivalence.
In this context, in view of their periods, the curves $E_+, {\cal E}_+$ are 2-fold unramified coverings of each other (the same
for $E_-, {\cal E}_-$), and their equations are obtained from one another by the classical
Landen transformation (see also Proposition \ref{Landen} for details and references).
Then  ${\cal E}_- \times {\cal E}_+$ is a 4-fold covering of $E_- \times E_+$ and vice versa.

We will use Proposition \ref{ell_case} in Subsection 4.2
in the analysis of a particular case of linearization
of the Kovalevskaya top on two elliptic curves.
\medskip

In the sequel we will call the 2-dimensional variety $\Prym(C/E)$ {\it generic}, if it does not contain elliptic curves.   

\paragraph{Dual pencils of genus $\le 3$ curves.} 
Following \cite{bw85} (see also \cite{avm88}), a generic torus
$\Prym (C/E)$ carries a line bundle $\cal L$ defining polarization (1,2) and $|{\cal L}|$ gives a one-parametric
family $\{K_\la\}, \la\in {\mathbb P}$ of algebraic curves $K_\la \subset \Prym(C/E) \subset \Jac(C)$. 

Geometrically, each curve $K_\la$ is the intersection $\Prym(C/E) \cap \Theta_{C,\la}$,
where $\Theta_{C,\la} \subset \Jac(C)$ is an appropriate translation of the theta-divisor $\Theta_C$,
the zero locus of the theta-function of $C$. More precisely, following \cite{Mum}, all $\Theta_{C,\la}$ are  
translations of $\Theta_C$ by vectors of $E\in \Jac(C)$.

A general curve of $K_\la$ is smooth of genus 3 and non-hyperelliptic.
The family $\{K_\la\}$ is, in fact,
a pencil of curves\footnote{One should stress that
$\{K_\la\}$ is a pencil only in $\Prym(C/E)$, the family of equivalent curves in ${\mathbb P}^2$ does not
form a pencil.} whose base locus consists of 4 points $q_1,\dots, q_4$.

Next, the extended involution $\sigma \, :\, \Jac(C)\to \Jac(C)$ acts on
$\Prym(C/E)\subset \Jac(C)$ as a reflection
having $q_1,\dots, q_4$ and the 4 half-periods of $E$ as fixed points. As was shown in \cite{bw85},
for a generic smooth curve $K_\la$, $\sigma(K_\la) =K_\la$, and $K_\la/\sigma$ is an elliptic curve ${\cal E}_\la$.
The latter is a member of a family $\{{\cal E}_\la\}$ of generally elliptic curves, which,
along with $K_\la$, forms a family of 2-fold coverings $K_\la \to {\cal E}_\la$, all of them ramified at
$q_1,\dots, q_4$.

Remarkably, all these coverings define the same Prym variety, denoted as $\Prym(K/{\cal E})$,
which is dual to $\Prym(C/E)$.
Then, as one may expect, the whole construction can be ''reflected'': there is a pencil of
(generally smooth and non-hyperelliptic) genus 3 curves
$$
C_\la \subset \Prym(K/{\cal E}) \subset \Jac(K), \qquad \la\in {\mathbb P}
$$
having 4 base points $Q_1,\dots,Q_4$, and a family of 2-fold coverings $C_\la \to E_\la$, all ramified at
$Q_1,\dots, Q_4$, and a general curve of $E_\la$ is elliptic.
All these coverings lead to the same Prym variety $\Prym(C/E)$ dual to $\Prym(K/{\cal E})$.
In this connection it is natural to refer to the pencils $\{C_\la \}, \{K_\la\}$ as {\it dual} pencils. 

A pencil $\{C_\la \}$ will be called generic if the corresponding variety $\Prym(C/E)$ is generic. 
In the case the dual pencil $\{K_\la\}$ and the variety $\Prym(K/{\cal E})$ are generic as well. 
\medskip

This beautiful purely algebraic geometric construction of Barth has been inspired by the work of
Haine \cite{hai83} on the complex geometry of the 4-dimensional integrable top.
Yet, it would not be complete without important observations
made by Horozov and van Moerbeke in \cite{hvm89} for the
pencil $\{C_\la\}$ related to the tori of the Kovalevskaya top, although they hold for generic pencils as well.
We summarize these results of \cite{hvm89} in form of the following theorem.

\begin{theorem} \label{fibers} 
A generic pencil $\{C_\la\}$ of curves of genus $\le 3$ in $\Prym^*(C/E)$ contains 6 hyperelliptic genus 3 curves
$\tilde{\cal H}_1, \dots, \tilde{\cal H}_6$,
which are 2-fold unramified coverings of genus 2 curves.
Among the latter there are only 3 non-equivalent curves, and they are
birationally equivalent to $\tilde\Gamma_1, \tilde\Gamma_2, \tilde\Gamma_3$ in Diagram \eqref{diagram}.

Next, $\{C_\la\}$ also contains 12 singular fibers $S_1,\dots, S_{12}$, which upon desingularization lead to curves of genus 2.
The set of these 12 curves consists of  3 distinct groups of four equivalent curves, defining $\Gamma_1, \Gamma_2, \Gamma_3$ in
\eqref{diagram}.

Respectively, the dual pencil $\{K_\la\}$ in $\Prym(C/E)$ contains 6 hyperelliptic genus 3 curves,
which are 2-fold coverings of curves equivalent to $\Gamma_1, \Gamma_2, \Gamma_3$,
and 12 singular curves whose regularizations are birationally equivalent to
$\tilde\Gamma_1, \tilde\Gamma_2, \tilde\Gamma_3$.
\end{theorem}

Thus, once a family of curves in ${\mathbb P}^2$, equivalent to the curves
$C_\la \subset \Prym^*(C/E)$, is known, one can calculate explicitly equations of all the genus 2 curves
$\Gamma_\alpha, \tilde\Gamma_\beta$ in Theorem \ref{main_th}.

For the Prym variety arising as the complex invariant torus of the classical Kovalevskaya top,
such a family of curves was found in \cite{hvm89},
although without any relation with the quotient spectral curve $C$ in \eqref{sp_kow}.

Apparently, for a generic covering $C\to E$ described in Theorem \ref{gen_cover}
there is no general recipe of reconstruction of the families $\{C_\la\}, \{K_\la\}$ in ${\mathbb P}^2$.

\paragraph{Remark 2.} In some special cases of covering $C\to E$ (see Subsection 4.2)
the dual pencil $K_\la\subset \Prym(C/E)$
may contain a reducible curve $K^*$ being a union of two elliptic curves, say ${\cal E}_-, {\cal E}_+$, and
$\Jac(K^*)$ is the direct product ${\cal E}_- \times {\cal E}_+$.
In the same case $C_\la\subset \Prym^*(C/E)$ may contain
a hyperelliptic genus 3 curve $C^*$, which is a 2-fold unramified
covering of a genus 2 curve $G$, which, in turn, is 2-fold ramified covering of ${\cal E}_-, {\cal E}_+$.

There is a natural conjecture that this case corresponds to the situation described in
Proposition \ref{ell_case}, when the Jacobians of some of related genus 2 curves are replaced by direct
products of elliptic curves, however we did not find a rigorous proof of this.
\medskip

Theorem \ref{fibers} can be accompanied with following well-known property (see, e.g., \cite{acgh84}):

\begin{proposition} \label{3->2}
Any hyperelliptic genus 3 curve which is also a 2-fold covering of an elliptic curve, can be written in form
\begin{equation} \label{hyp_3}
S\, :\; W^2=(Z^2-b_1)\cdots (Z^2-b_4)\quad \text{with distinct} \quad
b_1,\dots, b_4 \in{\mathbb C}\setminus \{0 \}.
\end{equation}
It has two obvious commuting involutions
$$
\sigma_1\, :\, (Z,W)\to (-Z,W), \qquad \sigma_2\, :\, (Z,W)\to (-Z,-W),
$$
the first one having 4 fixed points, the second one no such points.
Then $S/\sigma_1$ is the elliptic curve $E\, : \, \mu^2= (\la-b_1)\cdots (\la-b_4)$, whereas
$S/\sigma_2$ is the genus 2 curve \\ $\Gamma\, :\; w^2 = z(z-b_1)\cdots (z-b_4)$. $\Jac(\Gamma)$ is a 2-fold
unramified covering of $\Prym(S/E)$.
\end{proposition}

Note that the last statement is consistent with those of Theorems \ref{fibers},  \ref{main_th}.

\paragraph{Dual genus 3 curves.} The last necessary ingredient is a simple scheme which,
for a given genus 3 curve $C\in C_\la$, allows to obtain a curve $K\in K_\la$.
Let, as above, $C$ be a covering of the elliptic curve $E$ as described in Theorem \ref{gen_cover}.
Consider the tower of curves
$$\begin{CD}
 C @ > \pi >> E @ > \iota >> {\mathbb P} =\{x \},
\end{CD} $$
where $\pi, \iota$ denote the corresponding coverings with branch points $Q_1,\dots Q_4\in C$ and, respectively,
$P_1,\dots, P_4\in E$. Now let $K$ be another curve covering an elliptic curve ${\cal E}$
with the corresponding tower of 2-fold coverings
$$\begin{CD}
 K @ > \tilde\pi >> {\cal E} @ > \tilde\iota >> {\mathbb P} =\{x \},
\end{CD} $$
where $\tilde\pi, \tilde\iota$ are ramified at $\tilde Q_1,\dots \tilde Q_4\in K$, respectively,
$\tilde P_1,\dots, \tilde P_4\in {\cal E}$. 
We now require
\begin{equation} \label{flip}
  \tilde \iota \circ \tilde\pi (\tilde Q_i) = \iota(P_i), \quad \tilde\iota (\tilde P_i) =
\iota \circ \pi (Q_i), \qquad i=1,2,3,4,
\end{equation}
that is, $x$-coordinates of the branch loci of $C\to E$ (of $E\to {\mathbb P}$) coincide with
$x$-coordinates of the branch loci of  ${\cal E}\to {\mathbb P}$ (of $K\to {\cal E}$ ).


For a given generic covering $C\to E$, the above condition itself does not determine $K$ uniquely:
item 2) of Theorem \ref{gen_cover} says that 
there are four non-equivalent curves $K$ with the same branch points of $K\to {\cal E}$.
We then choose the curve $K$ in the form
\begin{gather} \label{dual}
 K\, :\; w^2 =g_3(x) + \rho(x) \sqrt{ (x-x_1)\cdots (x-x_4) } , \\ \qquad \text{or, equivalently}, \quad
  [ w^2- g_3(x)]^2 - \Psi(x) =0 , \notag
\end{gather}
with the polynomials $\rho(x), \Psi(x)$ specified in \eqref{branch_cond}. Observe that
$K$ is a 2-fold covering of the elliptic curve ${\cal E}\,: \; y^2=(x-x_1)\cdots (x-x_4)$, and the covering
is ramified at 4 points with $x=c_1,c_2,c_3, \infty$, which are the branch points of $E\to {\mathbb P}$.
Thus the conditions \eqref{flip} are satisfied.
A similar transformation applied to $K$ yields back the curve isomorphic to $C$ in \eqref{gen_gen_3}.

The following theorem is a reformulation of the result obtained in Section 3 of \cite{pan86}.

\begin{theorem} \label{bigonal}
For the curves $C,K$ described by \eqref{gen_gen_3}, \eqref{dual}, the corresponding varieties
$\Prym(C/E), \Prym(K/{\cal E})$ are dual. 
\end{theorem}

As a result, $C\subset \Prym(K/{\cal E}), \, K\subset \Prym(C/E)$ and, therefore, 
$C\in \{C_\la\}, K\in \{K_\la\}$.

It is natural to call the curves $C, K$ {\it dual}, or following \cite{pan86},
{\it bigonally related}, as the above construction is reminiscent (but not equivalent)
to Donagi's tetragonal construction involving unramified coverings of
curves\footnote{One should stress that $C$ and $K$ are not birationally equivalent.}.

\section{Curves of separation for the generalized Kovalevskaya gyrostat in one and two fields}
We now apply the above construction of dual pencils to the curves arising in the reductions 
of the Kovalevskaya gyrostat.

Recall that, according to Theorems \ref{Prym-tori}, \ref{Prym-tori2}, generic complex invariant manifolds of this system are open subsets of
$\Prym(C/E)$, where the quotient spectral curve $C$ and the underlying elliptic curve $E$
have the properties of Theorem \ref{gen_cover}. If the system is linearized on the Jacobian
of a genus 2 curve $G$ (a curve of separation of variables),
then $\Jac(G)$ is isogeneous to $\Prym(C/E)$.
And, if the isogeny is of degree 2, then, according to Theorem \ref{main_th},
$G$ must be equivalent to one of the genus 2 curves
$\Gamma_\alpha, \tilde\Gamma_\alpha$ in diagram \eqref{diagram}. We intend to calculate some
of these curves by using Theorem \ref{fibers} without finding the family
$C_\la\subset \Prym^*(C/E)$ explicitly.

\subsection{The Kovalevskaya gyrostat in one field}
As was already noticed, for a non-zero value of the area integral $I_1$,
the version $C_1$ of $C$ in \eqref{sp_kow} is smooth and
non-hyperelliptic (even for $\gamma =0$).
Observe however that, by the formula \eqref{dual}, the dual to $C_1$ reads
\begin{align} \label{kow_dual}
  K_1\, :\;  y^2 & = (x^2-(2H+\gamma^2) x+2) x + x \sqrt{R_4(x)}\, , \\
   & \qquad R_4=x (x^3 + (4 I_1-4H-2\gamma^2 )x^2 + (4 H \gamma^2+ I_2+\gamma^4)x -4\gamma^2) \,. \notag
\end{align}
For $\gamma \ne 0$, $K_1$ is again smooth, genus 3, and non-hyperelliptic, but for
$\gamma = 0$ (the case of the classical Kovalevskaya top) the polynomial $R_4(x)$, defining
the underlying elliptic curve $\cal E$, has a double root $x=0$,
and $K_1$ becomes singular of geometric genus 2.
By the birational transformation
$$
 z= \frac{y^2 -2x}{x^2}, \quad w = 2y \,\frac{2 I_1 x^2 -2x+y^2}{x^2}
$$
it is send to the hyperelliptic form
$$
 \hat {\cal K}\, :\;  w^2 = \left( (z+2 H)^2 - I_2 \right) \left( z( (z+ 2 H)^2 - I_2 +4 ) + 8 I_1 \right).
$$
Now one can easily observe that, under the substitution $z= - 2\xi$, $w= 4\sqrt{-2}\, \eta$, the above
equation takes the form of the original Kovalevskaya curve $\cal K$ in \eqref{KC}. As a result,
we conclude:
{\it Modulo birational transformations, the Kovalevskaya curve of separation of variables is dual to the quotient
genus 3 spectral curve of the Lax pair given in \cite{brs89}.}

Then, according to Theorems \ref{fibers}, \ref{main_th}, $\cal K$ is equivalent to one of the genus 2
curves $\widetilde\Gamma_\alpha$ in diagram \eqref{diagram}, and $\Jac({\cal K})$ is a 2-fold covering
of $\Prym(C/E)$, the compactified complex invariant manifold $\hat{\cal I}$ of the Kovalevskaya top. 
Equivalently,
$\hat{\cal I}$ is an 8-fold covering of the torus conformally equivalent to $\Jac({\cal K})$, as was
previously shown in \cite{hvm89} without using the spectral curve $C_1$.

\paragraph{The case $\gamma=0, I_1=0$.} The above configuration also holds for zero value of the
area integral: the Kovalevskaya curve becomes
$$
  {\cal K}_0\, :\; w^2 = s( (s-H)^2+1-I_2/4)((s-H)^2-I_2/4) ,
$$
it is still smooth of genus 2, and its Jacobian is a 2-fold covering of $\cal I$.

On the other hand, as was already noticed in \cite{brs89}, in this case
the quotient spectral curve $C_1$ is itself singular of genus 2 having the hyperelliptic form
$C_{10}$ in \eqref{C_10}. Then, Theorem \ref{fibers} implies that $C_{10}$ must be equivalent
to one of the curves $\Gamma_\alpha$ in \eqref{diagram}, and, by Theorems \ref{Prym-tori}, \ref{main_th},
$\Prym(C_1/E_1)\cong \hat{\cal I}$ is a 2-fold covering of $\Jac(C_{10})$.

Thus $\Jac({\cal K}_0)$ and $\Jac(C_{10})$ are isogeneous, one is a 4-fold covering of another, so
it is natural to expect that ${\cal K}, C_{10}$ are connected via the Richelot transformation.
And the authors of \cite{lm99} showed that this indeed holds by presenting this transformation explicitly.

In this connection it is also worth mentioning another genus 3 curve $B$
appearing in \cite{Kuz2} as the spectral
curve of a $2\times 2$ Lax pair for the Kovalevskaya--Goryachev--Chaplygin top, under the condition $I_1=0$.
As was shown in \cite{mark91}, $B$ can be transformed to the following hyperelliptic form in
variables $(u,v)$\footnote{The meaning of the constants is the same as above.}
$$
 B\,:\; v^2 = (u^4 - 2H u^2 +I_2/4) (u^4 - 2H u^2 +I_2/4-1) .
$$
This curve has the structure \eqref{hyp_3} in Proposition \ref{3->2},
hence it is 2-fold covering of an elliptic curve $E$ and of the genus 2 curve
$w^2=z(z^2 - 2H z +I_2/4) (z^2 - 2H z +I_2/4-1)$, the latter being precisely the curve $C_{10}$ in
\eqref{C_10}. Then $\Jac(C_{10})$ is a 2-fold covering of $\Prym(B/E)$ and the following diagram of
coverings holds (see also \cite{mark91})
 $$\begin{CD}
 \Jac({\cal K}_0) @ > 2:1 >> \Prym(C_1/E_1) @ > 2:1 >> \Jac( C_{10}) @ > 2:1 >> \Prym(B/E)   \,.
\end{CD} $$

\paragraph{The case $\gamma\ne 0, I_1=0$.} As in the previous case,
the curve $C_1$ in \eqref{sp_kow} is singular of genus 2
whose hyperelliptic form is a slight generalization of \eqref{C_10}, namely
$$
C_{10}'\, :\; Y^2= (X^2+2 H_g X+ \gamma^2 H_g +K_g/4-1) ( X^3+2 H_g X^2 + (K_g/4 + \gamma^2  H_g) X+\gamma^2 /2)\, .
$$
Here we set $H_g=H+\gamma^2 /2$, $K_g= I_2 -\gamma^4$. Then, following Theorem \ref{fibers},
$C_{10}'$ can be identified with one of the curves $\Gamma_\alpha$ in diagram \eqref{diagram}, and
the invariant manifold $\hat {\cal I}$ is a 2-fold covering of $\Jac(C_{10}')$.
Hence $C_{10}'$ is a curve of separation of variables
for the Kovalevskaya gyrostat on the level $I_1=0$.

In the same case, the dual curve \eqref{kow_dual} remains to be smooth of genus 3
and non-hyperelliptic, so one cannot
immediately apply Theorem \ref{fibers} to find equations of one of the curves in diagram \eqref{diagram}.
However, given a curve $\Gamma_\alpha$, to calculate one of the curves
$\widetilde\Gamma_\alpha$ one can make use of the
Richelot relations \eqref{tau->2tau}, that is, to apply the Richelot transformation to the known
curve $C_{10}'$. Omitting intermediate calculations, we present the result in the form
\begin{align}
{\cal K}_\gamma\, :\;  w^2 & = U_1(s) \, U_2(s)\, U_3(s), \label{Kgg} \\
 U_1 & = \rho s^2 + 2(1 + 2 H_g\rho -\rho^2)s +2H_g - \frac{\gamma^2 }{2}
+ \left(1 +4 H_g H - \frac{K_g}{2} \right)\rho -\rho^3,  \notag \\
 U_2 & = s^2 - 2\rho s - K_g/4 - \gamma^2  H_g +1 -2 H_g \rho, \notag \\
 U_3 & = s^2 - 2\rho s  - K_g/4 - \gamma^2  H_g - 4  H_g \rho - 2 \rho^2, \notag
\end{align}
where $\rho$ is the ''simplest'' root of the cubic polynomial in the equation of $C_{10}'$, 
\begin{gather*}
 \rho = \frac{\nu^{1/3}}{6} + \left( \frac 83 H_g^2 - 2\gamma^2  H_g- \frac{K_g}{2} \right) \frac{1}{\nu^{1/3}}
-\frac 23 H_g \, ,
\end{gather*}
and $\nu$ is a solution of the quadratic equation
$$
\nu^2 + (128 H_g^3+108 \gamma^2 -144 H_g (K_g/4+\gamma^2  H_g))\, \nu 
+ 4^3 (4 H_g^2-3 (K_g/4+\gamma^2  H_g))^2=0.
$$

It is a simple exercise to show that $\rho\to 0$ as $\gamma\to 0$ (and, therefore, $H_g\to H, K_g \to I_2$),
then the equation of ${\cal K}_\gamma$ reduces to $w^2 = 2 (s+H) (s^2+1 -K/4)(s^2-K/4)$,
which transforms to the Kovalevskaya curve ${\cal K}_0$
for $I_1=0$ by the birational change $w= \sqrt{2} \, \eta$, $s=\chi -H$.

Thus, due to Theorem \ref{main_th}, we conclude that the Jacobian of ${\cal K}_\gamma$ is a 2-fold
unramified covering of $\hat{\cal I} \cong \Prym(C_1/E_1)$, or, equivalently, the latter is an 8-fold covering of the complex
torus conformally equivalent to $\Jac({\cal K}_\gamma)$, hence the curve \eqref{Kgg} is a separation curve for the gyrostat on the level $I_1=0$.  This can be summarized in the diagram
 $$\begin{CD}
 \Jac({\cal K}_\gamma) @ > 2:1 >> \Prym(C_1/E_1) @ > 2:1 >> \Jac(\bar C_{10}') \,.
\end{CD} $$

The structure of ${\cal K}_\gamma$ suggests that in the construction of a separation of variables
for the Kovalevskaya gyrostat leading to a generalization of
the curve $\cal K$ for $\gamma\ne 0$, one cannot avoid solving algebraic equations of degree $\ge 3$.

There are numerous publications devoted to linearization of the Kovalevskaya gyrostat on the Jacobians
of genus 2 curves, both for $I_1=0$ and $I_1\ne 0$, see \cite{BorMam2, Kuz0, KZ04}, however we
did not find there any explicit expression for a separation genus 2 curve.

For various deformations of the gyrostat, the Lax pairs and separation curves have been presented in \cite{st02}.

\paragraph{The general case $\gamma\ne 0, I_1 \ne 0$.} As was already mentioned, in this case neither the original quotient
spectral curve $C_1$ not its dual $K_1$ are singular or hyperelliptic, hence one cannot apply Theorem \ref{fibers} to
obtain the corresponding genus 2 curves $\tilde\Gamma, \Gamma$ in diagram \eqref{diagram} without knowing explicitly
the pencil of genus 3 curves $C_\la \subset \Prym^*(C/E)$.

\subsection{The gyrostat in two fields} Here, as above, we consider only the case
$|g|=|h|=1, \langle g,h \rangle =0$, the corresponding quotient spectral curve
$C_2$ given in \eqref{sp_gen}. This situation is, in a sense, simpler than the
previous one: as Maple command {\tt Weierstrassform} shows, for generic constants of motion,
$C_2$ is hyperelliptic of genus 3 and is equivalent to
\begin{gather}
\hat C_2\, :\;  W^2 ={\cal R}(Z^2), \label{R8} \\
 {\cal R}(z)= z^{4} + 4 H_g\, z^{3} + (2 \gamma^2 \,H_g + \frac {1}{2} K_g- 16 + 4 H_g^2)\,z^2 \notag \\
 \quad + (4 \gamma^2 \,H_g^{2} + H_g\,K_g  + 8\,I_1 - 16\,H_g - 8\,\gamma^2  )\,z \notag  \\
 \qquad + \gamma^2 \,H_g K_g/2 + \frac {K_g^2}{16}+ 4\,I_1\,\gamma^2  - 8\,\gamma^2 \,H_g + \gamma^{4}\,H_g^2\, ,
\label{calR}
\end{gather}
where, as above, $H_g=H+\gamma^2 /2$, $K_g= I_2 -\gamma^4$.
The birational transformation between \eqref{R8} and \eqref{sp_gen} is described by
$$
 Z=\frac{y}{\sqrt{2}\, x}, \quad
 W= \frac{4}{x^4}\left( (K_g+4 \gamma^2  H_g)x^4-8\gamma^2  \,x^3+4 H_g y^2 x^2-8 xy^2+y^4\right) .
$$
Thus $\hat C_2$ has the structure of \eqref{hyp_3}, and, by Proposition \ref{3->2},
it is 2-fold covering of the elliptic curve $Y^2={\cal R}(z)$
(which is birationally equivalent to $E_2$ in \eqref{E22}, as expected)
and of the genus 2 curve $G =\{ w^2=z \,{\cal R}(z)\}$.  According to Theorems \ref{fibers}, \ref{main_th}, $G$ is
equivalent to one of the curves $\widetilde\Gamma_\alpha$ in diagram \eqref{diagram}, and the following holds.

\begin{proposition} \label{GG}
The complex invariant torus $\hat{\cal I}\cong \Prym(C_2/E_2)$ of the gyrostat in two fields is isogeneous to the Jacobian of
 $G =\{ w^2=z {\cal R}(z)\}$,  ${\cal R}(z)$ being specified in \eqref{calR}. More precisely, $\Jac(G)$ is a 2-fold
covering of $\Prym(C_2/E_2)$ or, equivalently, $\hat{\cal I}$ is an 8-fold covering of the torus conformally equivalent to
 $\Jac(G)$.
\end{proposition}

Note that under the shift $z= {\bf z}-H_g= {\bf z}- H-\gamma^2/2$, the equation of $G$ simplifies to 
\begin{equation} \label{G_simple}
G\, :\;  w^2 = ({\bf z}- H-\gamma^2/2) 
\left({\bf z}^4 - \left( 2 L+16 \right) {\bf z}^2 +(8 I_1+16H){\bf z} +L^2 -8 I_1 H \right) ,
\end{equation}
where we set $L=H^2-I_2/4$.

It follows that the curve $G$ can be a curve of
separation of variables for the reduced Kovalevskaya gyrostat in two fields with general constants of motion.
To our knowledge, an explicit linearization of this system is still an open problem.
There are however several papers presenting linearizations for particular cases of motion.

In Section 3 of \cite{Harl_Yehia87}, Harlamov and Yehia gave an explicit separation of variables
for the case $F_2=-2\gamma$, where, as above, $F_2=l_3-\langle l,[g,h] \rangle$ is the Hamiltonian of
the $SO(2)$-symmetry. Following \cite{Harl_Yehia87}, in this case
the reduced Lagrangian of the problem (the Routh function) does not contain linear terms in velocities.
Then the corresponding Hamilton function on $T^*S^2$ has a St\"ackel form, which ensures a 
separation of variables. 
The corresponding quadratures involve a genus 2 curve, which, for $|g|=|h|=1$,
{\it in our notation}, reads
\begin{gather} \label{HY_curve}
{\cal G}\, :\;  W^2 = (Z^2-\hat I_2-8\gamma^2) (Z^2-\hat I_2+ 8\gamma^2) (Z^2+2\gamma^2 Z- I_2 +\gamma^4)\, , \\
\hat I_2 = I_2+ 4\gamma^2 H+ \gamma^4. \notag
\end{gather}

Here one should mention that previously, in \cite{BorMam}, the Kovalevskaya gyrostat for $F_2=-2\gamma$ 
was shown to be isomorphic with Goryachev's particular case of integrability of the Kirchhoff equations
(\cite{Gor}), and the latter system had been reduced to quadratures, by different methods, in \cite{Zig0, Zig2, Ryabov}.  

Now, observe that, in view of  \eqref{I_F}, the condition $F_2=-2\gamma$ implies $I_1=2H$. 
Substituting this into \eqref{G_simple} and replacing ${\bf z}, w$ by ${\bf z}/2, w/2$, 
we obtain the following equation of the curve $G$
\begin{equation} \label{GY}
  G\, : \quad w^2 =( {\bf z}- 2 H_g) ({\bf z}^2 +8{\bf z} +I_2-16 H-4H^2) 
({\bf z}^2 - 8{\bf z}+I_2+ 16 H-4H^2) .
\end{equation}
Since $\Jac(G)$ is isogeneous to the invariant torus $\hat{\cal I}$, it can also be a
curve of separation of variables for $I_1=2H$.

An algorithm given in \cite{igusa62} allows to compare 
the absolute invariants of the curves \eqref{GY}, \eqref{HY_curve} in a numerical example, and the
result indicates that the curves are not birationally equivalent. 
On the other hand, the structure of \eqref{GY} is especially
convenient to apply the Richelot transformation to it, which yields the following curve
$$
   {\bar W}^2 = 16 ({\bar Z}^2 -4H \,{\bar Z} +4H^2 -K) ({\bar Z}^2- 4H_g {\bar Z} -8\gamma^2+4H^2 -I_2)
({\bar Z}^2- 4H_g {\bar Z} +8\gamma^2+4H^2 -I_2).
$$
Under the trivial substitutions $\bar Z=Z+2H_g, \bar W= 4W$, the latter takes the form \eqref{HY_curve}.

As a result, we have recovered the separation curve \eqref{HY_curve} of \cite{Harl_Yehia87} for the
case $F_2=-2\gamma$ from the quotient spectral curve $C_2$.

Due to the Richelot relation between the curves $G, {\cal G}$ and according to diagram \eqref{tau->2tau},
$\Jac(G)$ is a 4-fold covering $\Jac({\cal G})$. Then Theorem \ref{main_th} implies
the following tower of unramified coverings
 $$\begin{CD}
 \Jac(G) @ > 2:1 >> \Prym(C_2/E_2)= \hat{\cal I} @ > 2:1 >> \Jac({\cal G}) \, .
\end{CD} $$
\medskip

More curves of separation in particular cases of motion can be obtained by considering
 the curve dual to $C_2$ in \eqref{sp_gen}, which reads
\begin{equation} \label{dual_C2}
 K_2\, :\;  w^2 = (2 x^2-2H_g x+2) x + x \sqrt{ (4 I_1-8 H_g)x^3 + (K_g+2\gamma^2  H_g)x^2-4 \gamma^2  x  }\, ,
\end{equation}
and it is 2-fold covering of the generally elliptic curve
$$
{\cal E}_2\, :\;  y^2= (4 I_1-8 H_g)x^3 + (K_g+4\gamma^2  H_g)x^2-4 \gamma^2  x \, .
$$
For $\gamma\ne 0$ and general constants of motion, $K_2$ is smooth of genus 3 and,
in contrast to $C_2$, non-hyperelliptic.
It becomes singular of geometric genus 2 in two obvious cases: \\
1) $\gamma=0$ (${\cal E}_2$ is singular); \;
2) $I_1= 2 H_g \equiv 2H+ \gamma^2$ (${\cal E}_2$ is of genus 0).

In the first, i.e., gyroscopic-free, case, in view of \eqref{I_F}, there is the relation $I_1=2H+F_2^2$, and
the curve $K_2$ admits the following hyperelliptic form
\begin{equation} \label{K20}
K_{20}\,:\; w^2 = (z^2-I_2) \left[ z^4+ 2 F_2^2 z^3 -2 (2 H F_2^2 +I_2) z^2 -2I_2 F_2^2 z + I_2^2
+ 4  I_2 H F_2^2 + 16 F_2^4 \right] \, ,
\end{equation}
The corresponding birational transformation is
$$
 z= \frac{2x+2x^3 -y^2}{x^2}+2H, \quad w = 4 \sqrt{2}\, (I_1-2H)^{3/2}\, y .
$$
Thus, following Theorem \ref{fibers}, for $\gamma=0$,
$\Jac(K_{20})$ is a 2-fold covering of the invariant torus $\Prym(C_2/E_2)$.

\paragraph{Remark 3.} One can then compare $K_{20}$ with the curve $G$ in Proposition \ref{GG} for $\gamma=0$,
which takes the form
\begin{equation} \label{G0}
 G_0 \, :\;  w^2= z \left ( z^{4} + 4 H\, z^{3} + \left (\frac {I_2}{2} - 16 + 4 H^2 \right)\,z^2
 + (H I_2  + 8\, F_2^2)\,z +\frac {I_2^2}{16} \right) .
\end{equation}
Both $\Jac(G_0)$ and $\Jac(K_{20})$ are 2-fold coverings of $\Prym(C_2/E_2)$, however
the curves themselves are not birationally equivalent: in a numerical example
their Igusa absolute invariants are distinct.
Hence, $K_{20}, G_0$ represent two different curves $\widetilde\Gamma_\alpha$ in diagram
\eqref{diagram}, and both curves can serve as separation curves for the case $\gamma=0$.
\medskip

Observe that the second special case $I_1= 2 H_g$ is different from $I_1= 2 H$
considered in \cite{Harl_Yehia87} and above. Now the curve $K_2$ in \eqref{dual_C2} can be 
transformed to the following hyperelliptic form
\begin{equation} \label{K2p}
K_2'\, :\;  v^2 = (u^2+2\gamma^2 u -4\gamma^2 H- I_2 ) \,
\left[ (u^2-I_2 ) (u^2+2\gamma^2 u -4\gamma^2 H- I_2 )+16\gamma^4 \right] 
\end{equation}
with the transformation relations 
$$
 u= \frac{y^2-2x-2x^3}{x^2}+2H, \quad v= i \frac{8 \gamma^3 y}{\sqrt{2}\, x^2}. 
$$ 

It follows that $\Jac(K_{2}')$ is a 2-fold covering of the invariant torus $\Prym(C_2/E_2)$, and $K_2'$
is a possible curve of separation for $I_1= 2 H_g$. Apparently, an explicit separation of variables in this case is still an open problem.

\paragraph{The case $\gamma=0, F_2=0$: linearization on two elliptic curves.} 
As one can observe, in the limit
$\gamma=0, F_2=0$, which implies the relation $I_1=2H$, the equation of the
above curve $K_2'$ becomes $v^2=(u^2-I_2)^3$, so it is singular of genus 0. The same holds for the
curves ${\cal G}$ in \eqref{HY_curve} for $\gamma=0$ and $K_{20}$ in \eqref{K20} for $F_2=0$.

In this limit the original dual curve $K_2$ in \eqref{dual_C2} takes the algebraic form \\
$((2 x^2- 2Hx+2 )x-w^2 )^2 -I_2 x^4 =0$, which admits factorization
$$
   \left( (2 x^2- 2Hx+2 )x + \sqrt{I_2} x^2 - w^2 \right)\,
\left( (2 x^2- 2Hx+2 )x - \sqrt{I_2} x^2 - w^2 \right) =0 .
$$
It follows that $\Jac(K_2)$ is isomorphic to the direct product of two elliptic curves
\begin{equation} \label{EE}
 {\cal E}_\pm \, :\; w^2= 2 x (x^2 -(H \pm\kappa/2)x +1 ) , \qquad \kappa=\sqrt{I_2} \, ,
\end{equation}

This fact can be compared with the observation of \cite{BorMam, Yehia88} that in the case $F_2=0$
the two field Kovalevskaya top has a certain analogy with Chaplygin's particular integrable case of the
problem of motion of a body in a fluid \cite{Chapl}.
The latter case reduces to quadratures which consist of two different elliptic integrals. 
And in Section 5 of
\cite{Harl_Yehia87} a similar reduction was made for the Kovalevskaya top:
the corresponding quadratures involve the elliptic curves given by equations
\begin{equation} \label{Etau}
    T_1^2 = (\tau_1^2-1) (\tau_1-H/2+ \kappa/4 ), \quad  T_2^2 = (\tau_2^2-1) (\tau_2-H/2 - \kappa/4 ).
    \end{equation}

One can check however that these curves are not equivalent to ${\cal E}_\pm$ in \eqref{EE}.

On the other hand, observe that under the same conditions $\gamma=0, F_2=0$
the alternative genus 2 curve $G_0$ in \eqref{G0} remains smooth and takes the form
$$
G_{00}\,:\; w^2 =\frac{1}{16} z\, \left( z^2+4(H+2)z+ I_2 \right)\left( z^2+4(H-2)z+I_2 \right) .
$$
Then Theorem \ref{fibers} implies that $\Jac(G_{00})$ is a 2-fold unramified covering of the
complex invariant manifold $\hat {\cal I}$.

Upon setting again $\kappa=\sqrt{I_2}$ the curve reads
\begin{gather} \label{split}
   G_{00}\,:\; 16 w^2 =z (z-r_1 )(z-r_2) (z-\rho_1)(z-\rho_2), \\
   \begin{aligned}
     r_{1,2} &= -2H-4 \pm \sqrt{ (2H+4+\kappa) (2H+4-\kappa) },  \\
   \rho_{1,2} &= -2H+4 \pm \sqrt{ (2H-4+\kappa) (2H-4-\kappa) }. \end{aligned} \label{rr}
\end{gather}
One immediately sees that $r_1 r_2=\rho_1 \rho_2=I_2$. This case corresponds to the simplest possible
reduction of curves and Abelian functions studied by Jacobi and described in many books,
see, e.g., \cite{Bel, kra903}.

\begin{proposition}[\cite{Bel}] \label{Jacobi} Under the condition $r_1 r_2=\rho_1 \rho_2$,
the curve of the form \eqref{split} is a 2-fold ramified covering of two generally different
elliptic curves\footnote{Explicit formulas of the coverings are not used in this paper, so we do not give them.}
$$
E_\pm\, : \;  \mu^2 = \la (\la-1)(\la- \delta_\pm ), \quad
\delta_\pm = - \frac{\rho_1 ( \sqrt{r_1} \pm \sqrt{r_2})^2 }{ (r_1-\rho_1)(r_2-\rho_1) } .
$$
The Jacobian of $G_{00}$ contains $E_-, E_+$ as Abelian subvarieties, which intersect at 4 points.
$\Jac(G_{00})$ is not the direct product $E_+\times E_-$, but isogeneous to it:
$\begin{CD}
 E_+\times E_- @ > 4:1 >> \Jac(G_{00})
\end{CD}$.
\end{proposition}

In view of \eqref{rr}, $(r_1-\rho_1)(r_2-\rho_1) = 16 \rho_1$, hence in our case
$\delta_\pm = (2H+4\mp \kappa)/8$, and, under the change $\la=(X+1)/2$, $\mu=Y/\sqrt{8}$, the curves $E_\pm$ take the form
\begin{equation} \label{Epm}
E_\pm\, : \;  Y^2 =  (X^2-1)\left(X- \frac{H}{2} \pm \frac{\kappa}{4} \right).
\end{equation}
The latter coincide with the separation curves \eqref{Etau} obtained in \cite{Harl_Yehia87}
for the case $\gamma=0, F_2=0$.

The appearance of the direct product of curves ${\cal E}_\pm$, the pair $E_\pm$, and of the double coverings
 $\begin{CD}  E_+   @ < << G_{00} @ > >> E_-\end{CD}$ implies that the considered case is described
 by Proposition \ref{ell_case} and diagrams \eqref{special_Jacs}: \\
$\Jac(G_{00})$ and the product ${\cal E}_+ \times {\cal E_-}$
are (different) 2-fold unramified coverings of the complex invariant torus $\hat {\cal I}$, which, itself,
is a 2-fold unramified covering of the direct product $E_+ \times E_-$.

Then, following Proposition \ref{ell_case}, the elliptic curves $E_\pm$ and
${\cal E}_\pm$ must be isogeneous. And, indeed, ${\cal E}_+$  (respectively, ${\cal E}_-$) is obtained from $E_+$(respectively, $E_-$) by duplicating one of the two period vectors. The equations of the corresponding curves are obtained from each other by a second order transformation (also called {\it Landen's transformation}, see, e.g., \cite{acgh84}), which can be described as follows.

\begin{proposition}[\cite{acgh84}] \label{Landen}
The equation of any double unramified covering of an elliptic curve
$E =\{ \mu^2=(\la-a_1)(\la-a_2)(\la-a_3) \}$ is given by regularization $\hat E$ of the curve
$$
  \left\{ (\la,\mu,z)\in E\times {\mathbb P} \mid \mu^2=(\la-a_1)(\la-a_2)(\la-a_3) ,\; z^2 =\frac{\la-a_j}{\la-a_i} \;
\textup{or}\quad z^2= \la -a_i \right\} ,
$$
$i,j=1,2,3, \, i\ne j$. For example,
\begin{gather*}
\hat E=\left\{w^2= \left( z-\frac{a_i-a_k }{a_j-a_k} \right) (z^2-1) \right\}, \quad
\text{or, respectively,} \\
\hat E= \{w^2= (z^2+a_i-a_j) (z^2+a_i-a_k)\}, \quad (i,j,k)=(1,2,3) .
\end{gather*}
The covering $\hat E\to E$ is obtained by doubling the period $\int_{a_i}^{a_j} \frac{d\la}{\mu}$
(respectively, $\int_{\infty}^{a_i} \frac{d\la}{\mu}$) in the parallelogram of periods of $E$.
\end{proposition}

Following this procedure, we make the second order substitution
$$
   Z^2 = \frac{X-1}{X+1} \, \Longleftrightarrow \, X= \frac{1+Z^2}{1-Z^2}
$$
in the equation \eqref{Epm} of $E_+$ obtaining
$$
   Y^2 = - \frac{4 Z^2}{(Z^2-1)^4}\cdot (Z^2-1) \left[ \left( 1+\frac H2-\frac{\kappa}{4} \right)Z^2
+ 1 -\frac H2 + \frac{\kappa}{4} \right] .
$$
Regularizing this curve by omitting full squares, we get the following equation of the 2-fold covering of $E_+$
$$
   Y^2 = -  (Z^2-1) \left[ \left( 1+\frac H2-\frac{\kappa}{4} \right)Z^2 + 1 -\frac H2 + \frac{\kappa}{4} \right] .
$$
Then the birational substitution $Z=2/(z+1)-1$, $Y=\sqrt{-2} \, 4 w/(z+1)^2$ takes the above equation precisely to the curve
${\cal E}_+$ in \eqref{EE}. The same composition of changes takes $E_-$ to ${\cal E}_-$.

As a result, the product ${\cal E}_+ \times {\cal E_-}$  can be regarded as a 4-fold covering of
$E_-\times E_+$, and vice versa (if we identify conformally equivalent Abelian tori).
And, as stated in Proposition \ref{Jacobi}, $E_-\times E_+$ is also a 4-fold covering of $\Jac(G_{00})$. All the observations lead to the following diagram of coverings of Abelian varieties in the case $\gamma=0, F_2=0$. 

\begin{figure}[H]
\begin{center}
\includegraphics[width=0.45\textwidth]{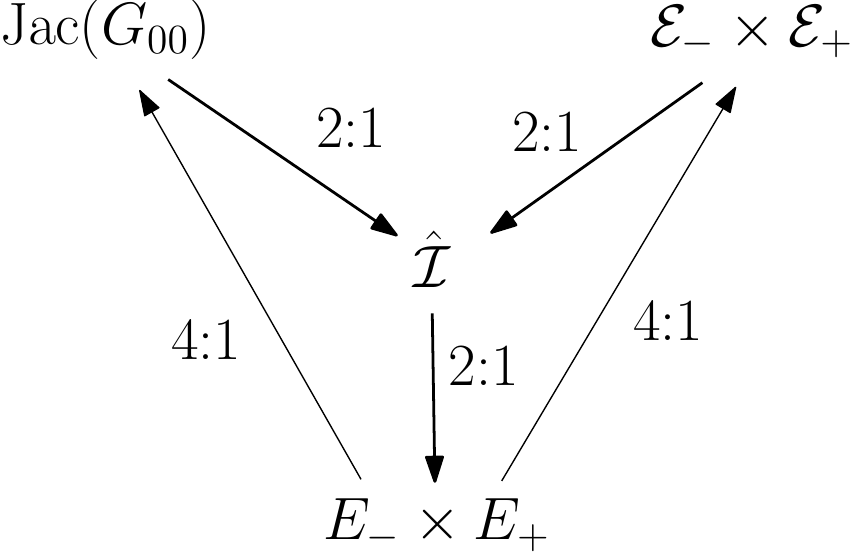}
\end{center}
\end{figure}

\vspace{ -10 pt}
According to Proposition \ref{ell_case}, apart from the curves $E_\pm, {\cal E}_\pm$, and $G_{00}$
there exist three other genus 2 curves: a curve $G_{00}'$, which is also a 2-fold covering of $E_-, E_+$, and two non-equivalent curves
$\Gamma_1, \Gamma_2$, being such coverings of ${\cal E}_-, {\cal E}_+$ (we do not present their equations here). 
All of them can be used as curves of separation in the considered
special case, as their Jacobians are isogeneous to the complex invariant torus $\hat {\cal I}$.    
\medskip

We conclude by summarizing all the curves of this subsection, as well as the relations between them,
in the following diagram where the arrows denote reductions to particular cases of motion or coverings, or the Richelot transformation,
and the symbol $\ncong$ indicates non-equivalence.

\begin{figure}[H]
\begin{center}
\includegraphics[width=0.8\textwidth]{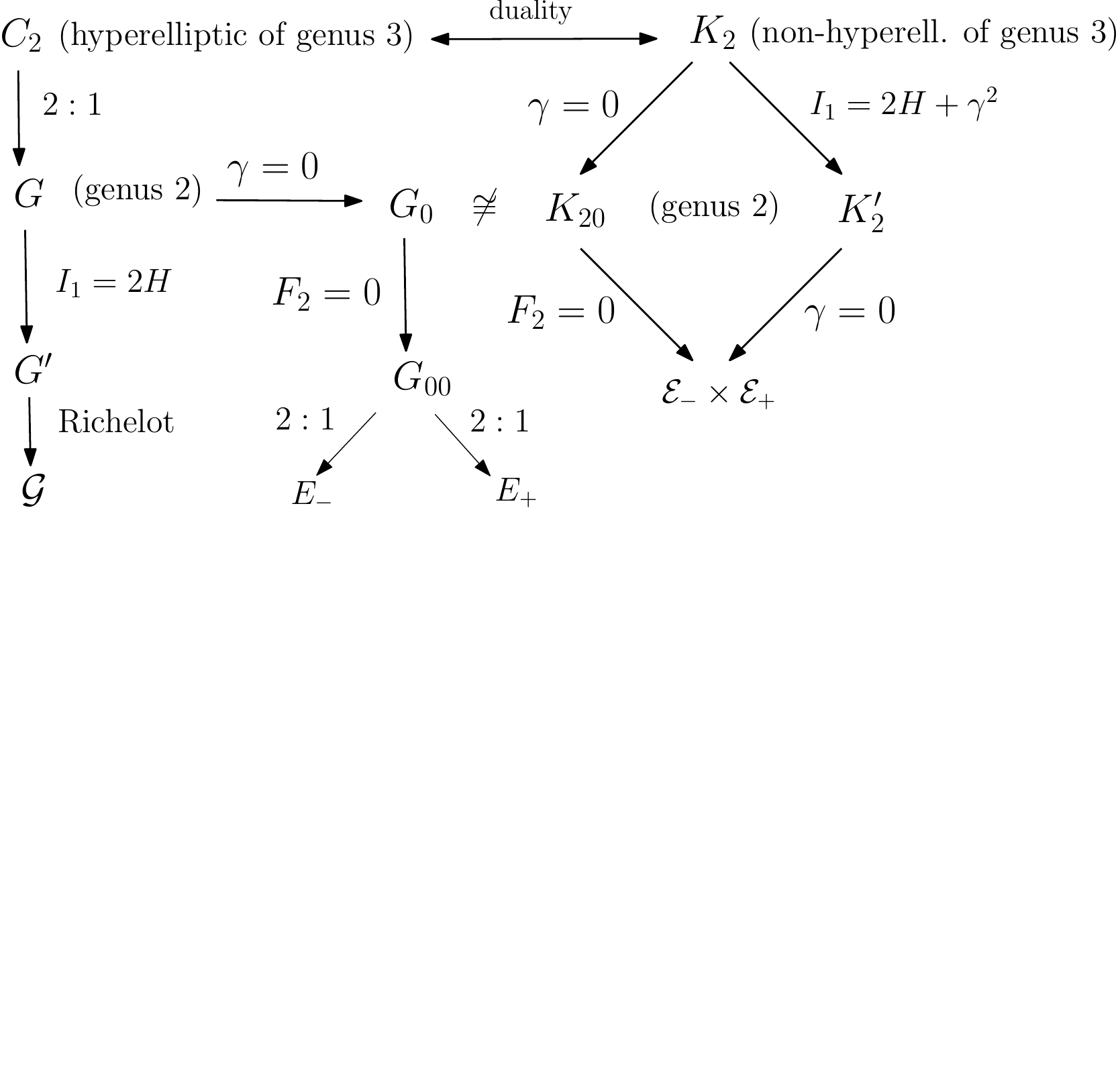}
\end{center} 
\end{figure}

\vspace{ -190 pt}

\section*{Conclusion} From the spectral curve of the Lax pair of the Kovalevskaya gyrostat in one and two fields presented
in \cite{brs89} we managed to recover all known curves of separation of variables, including the original
Kovalevskaya curve, and constructed new ones.

Naturally, the approach  we used does not allow to make an explicit reduction of the systems to quadratures. 
However, the obtained separation
curves can be used in analysis of bifurcations of complex invariant manifolds of the systems. 

Next, the approach itself does not permit to reconstruct explicitly the whole pencil of genus 3 curves 
$C_\la \subset \Prym^*(C/E)$, which might be used to derive alternative genus 2 curves, 
whose Jacobians are degree 2 isogeneous to $\Prym(C/E)$, and which are depicted in diagram \eqref{diagram}. 

Returning to the case of general configuration of the Kovalevskaya gyrostat in two fields, when
the conditions $|g|=|h|=1, \langle g,h \rangle =0$ do not hold, it is natural to conjecture that
generic complexified invariant tori of the system are isomorphic to the 3-dimensional Prym variety $\cal P$ 
associated with
the quotient genus 4 curve $C$ in \eqref{gen_C} covering an elliptic curve. 
Then an interesting open problem is to describe
principally polarized Abelian tori isogeneous to $\cal P$ as Jacobians of genus 3 curves. The latter will 
serve as curves of separation of variables for this integrable system.  

\section*{Acknowledgments} The authors are grateful to A. Borisov and P. Ryabov for 
discussions and clarifying remarks.

The contribution of Y.F was partially supported by the Spanish MINECO-FEDER Grants MTM2012-31714,  MTM2012-37070. He also gratefully acknowledges
the hospitality and a financial support of UNAM, Mexico City, where a part of this work has been completed. 
LGN acknowledges support for his research from the Program UNAM-DGAPA-PAPIIT-IA103815.

\end{document}